\documentclass[iop]{emulateapj}
\usepackage{amssymb}
\usepackage{apjfonts}
\usepackage{graphicx}

\usepackage{amsmath}
\usepackage{epstopdf}
\usepackage{natbib} 
\usepackage{lineno}
\usepackage{color}

\slugcomment{prepared for ApJ}
\begin{document}
\title{Galactic Cosmic-Ray Anisotropy in the Northern hemisphere from the ARGO-YBJ Experiment during 2008-2012}
\author{ B.~Bartoli\altaffilmark{1,2}, P.~Bernardini\altaffilmark{3,4}, X. J.~Bi\altaffilmark{5,6}, Z.~Cao\altaffilmark{5,6},
 S.~Catalanotti\altaffilmark{1,2}, S. Z.~Chen\altaffilmark{5}, T. L.~Chen\altaffilmark{7}, S. W.~Cui\altaffilmark{8},
 B. Z.~Dai\altaffilmark{9}, A. D'Amone\altaffilmark{3,4}, ~Danzengluobu\altaffilmark{7}, I.~De Mitri\altaffilmark{23,4},
 B. D'Ettorre Piazzoli\altaffilmark{1,10}, T.~Di Girolamo\altaffilmark{1,2}, G.~Di Sciascio\altaffilmark{10}, C. F.~Feng\altaffilmark{11},
 Z. Y.~Feng\altaffilmark{12}, W.~Gao\altaffilmark{5,6}, Q. B.~Gou\altaffilmark{5},
 Y. Q.~Guo\altaffilmark{5}, H. H.~He\altaffilmark{5,6}, Haibing~Hu\altaffilmark{7}, Hongbo~Hu\altaffilmark{5},
 M.~Iacovacci\altaffilmark{1,2}, R.~Iuppa\altaffilmark{13,14}, H. Y.~Jia\altaffilmark{11}, ~Labaciren\altaffilmark{7},
 H. J.~Li\altaffilmark{7}, C.~Liu\altaffilmark{5}, J.~Liu\altaffilmark{9}, M. Y.~Liu\altaffilmark{7},
 H.~Lu\altaffilmark{5}, L. L.~Ma\altaffilmark{5}, X. H.~Ma\altaffilmark{5}, G.~Mancarella\altaffilmark{3,4},
 S. M.~Mari\altaffilmark{15,16}, G.~Marsella\altaffilmark{3,4}, S.~Mastroianni\altaffilmark{2}, P.~Montini\altaffilmark{17},
 C. C.~Ning\altaffilmark{7}, L.~Perrone\altaffilmark{3,4}, P.~Pistilli\altaffilmark{15,16},
 D.~Ruffolo\altaffilmark{24},
 P.~Salvini\altaffilmark{18},
 R.~Santonico\altaffilmark{9,19}, P. R.~Shen\altaffilmark{5}, X. D.~Sheng\altaffilmark{5}, F.~Shi\altaffilmark{5},
 A.~Surdo\altaffilmark{4}, Y. H.~Tan\altaffilmark{5}, P.~Vallania\altaffilmark{20,21}, S.~Vernetto\altaffilmark{20,21},
 C.~Vigorito\altaffilmark{21,22}, H.~Wang\altaffilmark{5}, C. Y.~Wu\altaffilmark{5}, H. R.~Wu\altaffilmark{5},
 L.~Xue\altaffilmark{11}, Q. Y.~Yang\altaffilmark{9}, X. C.~Yang\altaffilmark{9}, Z. G.~Yao\altaffilmark{5},
 A. F.~Yuan\altaffilmark{7}, M.~Zha\altaffilmark{5}, H. M.~Zhang\altaffilmark{5}, L.~Zhang\altaffilmark{9},
 X. Y.~Zhang\altaffilmark{11}, Y.~Zhang\altaffilmark{5}, J.~Zhao\altaffilmark{5}, ~Zhaxiciren\altaffilmark{7},
 ~Zhaxisangzhu\altaffilmark{7}, X. X.~Zhou\altaffilmark{12}, F. R.~Zhu\altaffilmark{12}, and Q. Q.~Zhu\altaffilmark{5}\\
 (The ARGO-YBJ Collaboration)}

\altaffiltext{0}{Corresponding authors: Wei Gao, \textbf{gaowei@ihep.ac.cn}; Songzhan Chen, \textbf{chensz@ihep.ac.cn}; Huihai He, \textbf{hhh@ihep.ac.cn}}
\altaffiltext{1}{Dipartimento di Fisica dell'Universit\'{a} di Napoli "Federico II," Complesso Universitario di Monte Sant'Angelo, via Cinthia, I-80126 Napoli, Italy}
\altaffiltext{2}{Istituto Nazionale di Fisica Nucleare, Sezione di Napoli, Complesso Universitario di Monte Sant'Angelo, via Cinthia, I-80126 Napoli, Italy}
\altaffiltext{3}{Dipartimento Matematica e Fisica "Ennio De Giorgi," Universit\'{a} del Salento, via per Arnesano, I-73100 Lecce, Italy}
\altaffiltext{4}{Istituto Nazionale di Fisica Nucleare, Sezione di Lecce, via per Arnesano, I-73100 Lecce, Italy}
\altaffiltext{5}{Key Laboratory of Particle Astrophysics, Institute of High Energy Physics, Chinese Academy of Sciences, P.O. Box 918, 100049 Beijing, China}
\altaffiltext{6}{School of Physical Sciences, University of Chinese Academy of Sciences, 100049 Beijing, P.R.China}
\altaffiltext{7}{Tibet University, 850000 Lhasa, Xizang, China}
\altaffiltext{8}{Hebei Normal University, 050024 Shijiazhuang Hebei, China}
\altaffiltext{9}{Yunnan University, 2 North Cuihu Road, 650091 Kunming, Yunnan, China}
\altaffiltext{10}{Istituto Nazionale di Fisica Nucleare, Sezione di Roma Tor Vergata, via della Ricerca Scientifica 1, I-00133 Roma, Italy}
\altaffiltext{11}{Shandong University, 250100 Jinan, Shandong, China}
\altaffiltext{12}{Southwest Jiaotong University, 610031 Chengdu, Sichuan, China}
\altaffiltext{13}{Dipartimento di Fisica dell'Universit\'{a} di Trento, via Sommarive 14, I-38123 Povo, Italy}
\altaffiltext{14}{Trento Institute for Fundamental Physics and Applications, via Sommarive 14, I-38123 Povo, Italy}
\altaffiltext{15}{Dipartimento di Fisica dell'Universit\'{a} "Roma Tre," via della Vasca Navale 84, I-00146 Roma, Italy}
\altaffiltext{16}{Istituto Nazionale di Fisica Nucleare, Sezione di Roma Tre, via della Vasca Navale 84, I-00146 Roma, Italy}
\altaffiltext{17}{Dipartimento di Fisica dell'Universit\'{a} di Roma "La Sapienza" and INFN¡ªSezione di Roma, piazzale Aldo Moro 2, I-00185 Roma, Italy}
\altaffiltext{18}{Istituto Nazionale di Fisica Nucleare, Sezione di Pavia, via Bassi 6, I-27100 Pavia, Italy}
\altaffiltext{19}{Dipartimento di Fisica dell'Universit'\'{a} di Roma "Tor Vergata," via della Ricerca Scientifica 1, I-00133 Roma, Italy}
\altaffiltext{20}{Osservatorio Astrofisico di Torino dell'Istituto Nazionale di Astrofisica, via P. Giuria 1, I-10125 Torino, Italy}
\altaffiltext{21}{Istituto Nazionale di Fisica Nucleare, Sezione di Torino, via P. Giuria 1, I-10125 Torino, Italy}
\altaffiltext{22}{Dipartimento di Fisica dell'Universit\'{a} di Torino, via P. Giuria 1, I-10125 Torino, Italy}
\altaffiltext{23}{Gran Sasso Science Institute (GSSI), Via M. Iacobucci 2, I-67100 L'Aquila, Italy}
\altaffiltext{24}{Department of Physics, Faculty of Science, Mahidol University, Bangkok 10400, Thailand}
\begin{abstract}
This paper reports on the observation of the sidereal large-scale anisotropy of cosmic rays using data collected by the ARGO-YBJ experiment over 5 years (2008$-$2012). This analysis extends
previous work limited to the period from 2008 January to 2009 December,
near the minimum of solar activity between cycles 23 and 24.
With the new data sample the period of solar cycle 24 from near minimum to
maximum is investigated. A new method is used to improve the energy reconstruction, allowing us to cover a much wider energy range, from 4 to 520 TeV.
Below 100 TeV, the anisotropy is dominated by two wide regions, the so-called "tail-in" and "loss-cone" features. At higher energies, a dramatic change of the morphology is confirmed.
The yearly time dependence of the anisotropy is investigated.
Finally, no noticeable variation of cosmic-ray anisotropy with solar activity is observed for a median energy of 7 TeV.
\end{abstract}
\section{Introduction}
In the past decades, the sidereal directional variation of the Galactic cosmic-ray intensity, which we refer to as the anisotropy, was observed by many detectors across a wide energy range from 60 GeV to 8 EeV. The  morphology of anisotropy  and the corresponding amplitude are energy-dependent. The anisotropy implies important information about the magnetic structure of the heliosphere, the local interstellar space surrounding the heliosphere, and large portions of the Galaxy through which cosmic rays propagate to the Earth. The study of anisotropy can shed new light on the origin and propagation of cosmic rays.

Below 100 TeV, two large-scale features recognized as "tail-in" and "loss-cone" features in two dimensional (2D) maps are observed with very high significance both in the northern hemisphere \citep[]{AS2006, argo2015} and southern hemisphere \citep[]{icecube2016}. According to the ARGO-YBJ observations at energies from 1 TeV to 30 TeV, the amplitude of anisotropy increases with energy, reaching a maximum at around 10 TeV, while the angular phase is approximately stable \citep[]{argo2015}. Different models have been proposed to explain the origin of the anisotropy, concerning different aspects of cosmic-ray physics, from the sources of cosmic rays to the propagation to the Earth. Some models consider the anisotropy due to the spatial distribution of cosmic-ray sources, as the presence of a nearby strong source \citep[]{erly2006,liuwei2017}. Other interpretations concern the structure of the Galactic \citep[]{qu2012}, local interstellar \citep[]{schw2014} and interplanetary magnetic fields \citep[]{Nagashima1998}. Besides the large-scale anisotropy (LSA), some works focusing on medium scale anisotropy were also reported by Milagro \citep[]{milagro2008} and ARGO-YBJ \citep[]{barto13b}.

No variations with time are expected if the LSA is produced by a nearby source or due to the interstellar magnetic field, while a significant heliospheric influence would show time variation in association with the 11-year solar cycle. Several long term experiments allow time-dependent studies of the LSA; however, contradictory results were obtained. In the northern hemisphere, Milagro reported a steady increase in the amplitude of the "loss-cone" at 6 TeV over a seven-year time period (2000$-$2007) as the solar activity varied from near maximum to minimum \citep[]{milagro2009}. If this observation is true, it will be a challenge to find a consistent explanation of the observed anisotropy and the corresponding time evolution. However, no significant time variations were observed by Tibet AS$\gamma$  at 4.4, 6.2, and 11 TeV over a period that overlapped with the Milagro observation time, from 1999 to 2008 \citep[]{AS2012}. In the southern hemisphere, no time dependence was observed by IceCube at energies above 13 TeV over a six-year time period (2009-2015) as the solar activity varied from near minimum to maximum \citep[]{icecube2016}.

As the energy increases above 100 TeV, a major change in the morphology of the anisotropy is observed by several experiments. For such high energy cosmic rays, the influence of the magnetic field within the heliosphere on the cosmic-ray anisotropy is expected negligible. Therefore, the morphology should be time independent. The EAS-TOP collaboration reported a detection of a new anisotropy pattern at 370 TeV with a limited significance \citep[]{eastop2009} for the first time. Around 400 TeV, IceCube found that the structure of the anisotropy is mostly characterized by a wide relative deficit at a right ascension of 30$^{\circ}$-120$^{\circ}$ \citep[]{icecube2012}. With the accumulation of six years of data, the IceCube collaboration found that the change in the morphology starts from about 100 TeV and the amplitude of the deficit increases with energy up to at least 5 PeV \citep[]{icecube2016}. Recently, the Tibet AS$\gamma$ collaboration improved their analysis with detecting LSA at 300 TeV which is in close agreement with IceCube's results \citep[]{AS2017}. The study of LSA at energies above 100 TeV is helpful to understand the origin and propagation of Galactic cosmic rays. The observed phase of the excess region includes the direction of the Galactic center \citep[]{icecube2016,AS2017}, perhaps indicating the dominant source(s) of the cosmic rays \citep[]{guo2013a,guo2013b}. It is worth noting that the Auger collaboration have observed LSA at an energy above 8 EeV \citep[]{auger2017}. The morphology is different from that at 100 TeV$-$5 PeV. The direction of the anisotropy indicates an extragalactic origin for these ultrahigh energy particles.

Based on the first two years of data (2008 and 2009), the ARGO-YBJ collaboration have reported an energy dependence of the LSA at energies from 1 TeV to 30 TeV \citep[]{argo2015}. In this paper, we present a new analysis with five years data of the ARGO-YBJ experiment collected from 2008 to 2012. During this period, the solar activity varied from near minimum to maximum. Therefore, the time dependence of the anisotropy is analyzed to address the contradiction between Milagro and Tibet AS$\gamma$ data. Here, we also adopt an improved energy estimation to extend the maximum energy to above 100 TeV,
for which few results have been reported.

\section{Data and analysis methods}
\subsection{The ARGO-YBJ experiment and data}

The ARGO-YBJ experiment, located at the Yangbajing Cosmic Ray Laboratory (Tibet, China, 30.11 N, 90.53 E, altitude of 4300 m  a.s.l.), is mainly devoted to $\gamma$-ray astronomy
\citep[]{barto13,barto14,barto16}
and cosmic-ray physics
\citep[]{barto12,barto15b}.
The detector consists of a carpet ($\sim$74$\times$ 78 m$^2$) of resistive plate chambers (RPCs) with $\sim$93\% of active area, surrounded by a partially instrumented area ($\sim$20\%) up to $\sim$100$\times$110 m$^2$.
Each RPC is read out via 80 strips (6.75 cm$\times$61.8 cm), logically organized in 10 pads (55.6 cm$\times$61.8 cm).
The pad is the basic unit for timing and trigger purposes. Each pad can count the number of particles, up to 8.
More details of the detector and its performance can be found in \citet{aielli06,aielli09}.
The ARGO-YBJ experiment in its final configuration started taking data in 2007 November and stopped in 2013 January.
The ARGO-YBJ detector is operated by requiring the coincidence of at least 20 fired pads within 420 ns on the central carpet detector.
The trigger rate is 3.5 kHz with a dead time of 4\%. Through the whole operation period, the average duty-cycle is higher than $86\%$.
The data collected from January 2008 to December 2012 are used in this analysis.

\begin{figure*}
\includegraphics[width=0.32\textwidth]{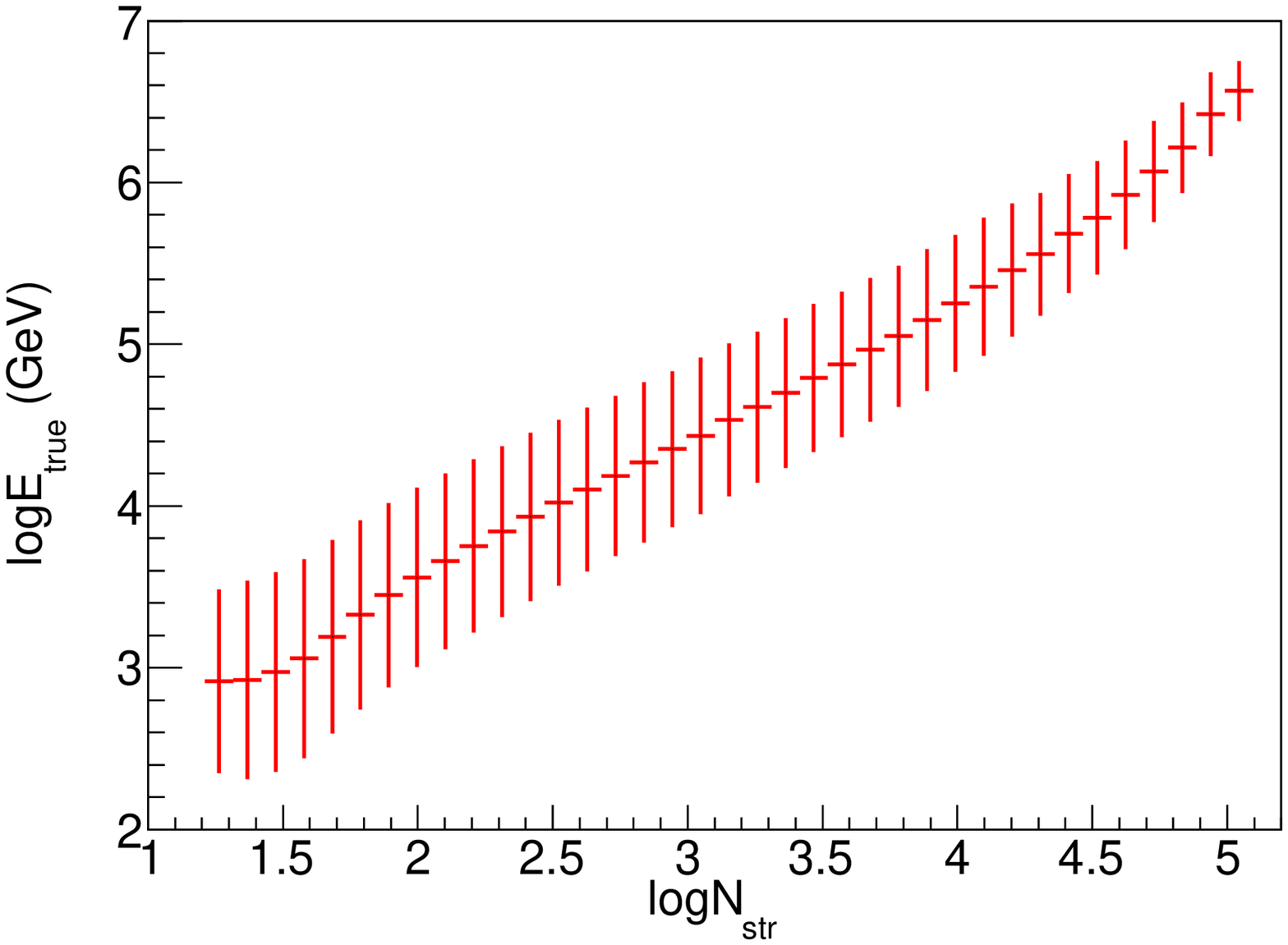}
\includegraphics[width=0.32\textwidth]{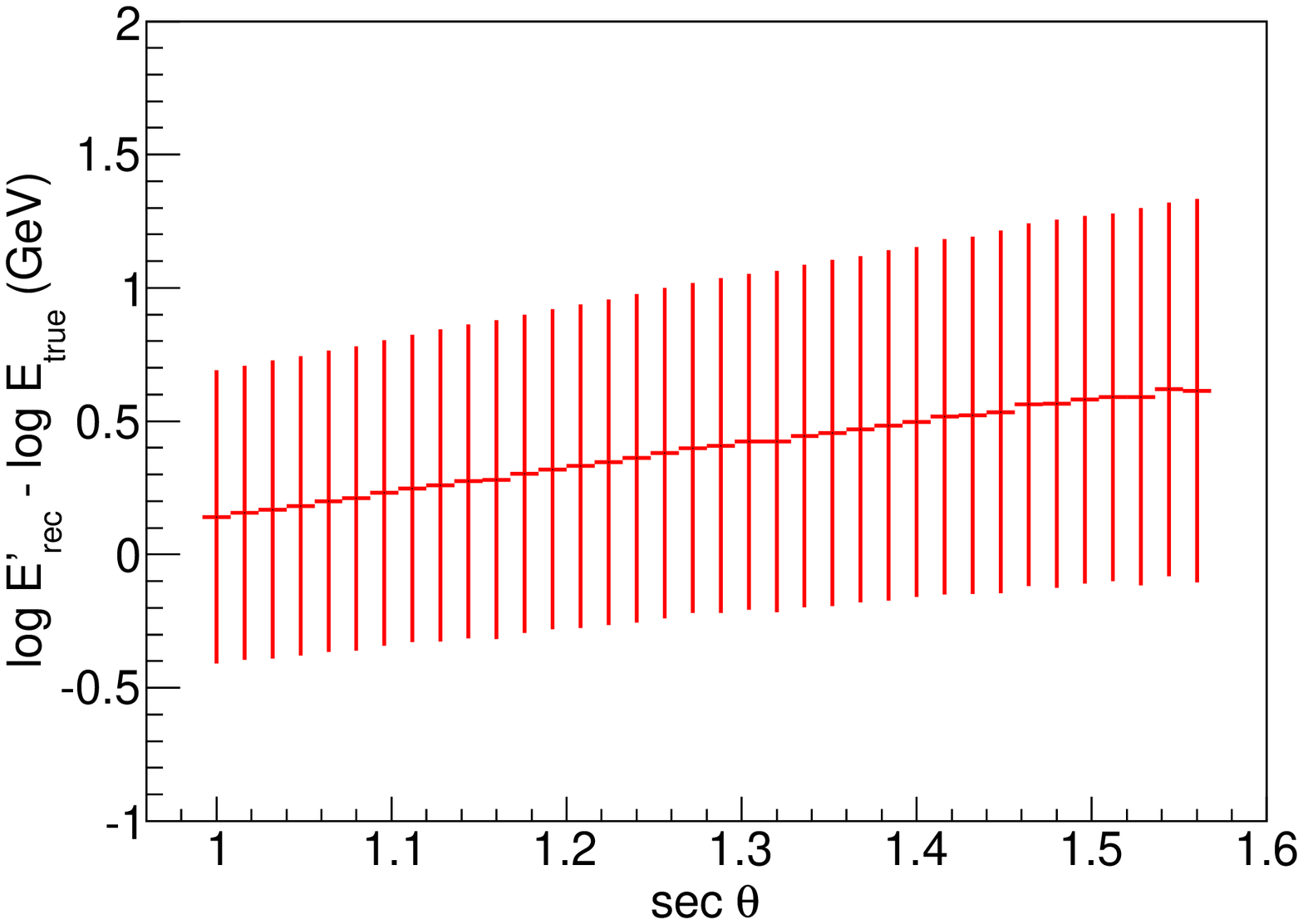}
\includegraphics[width=0.32\textwidth]{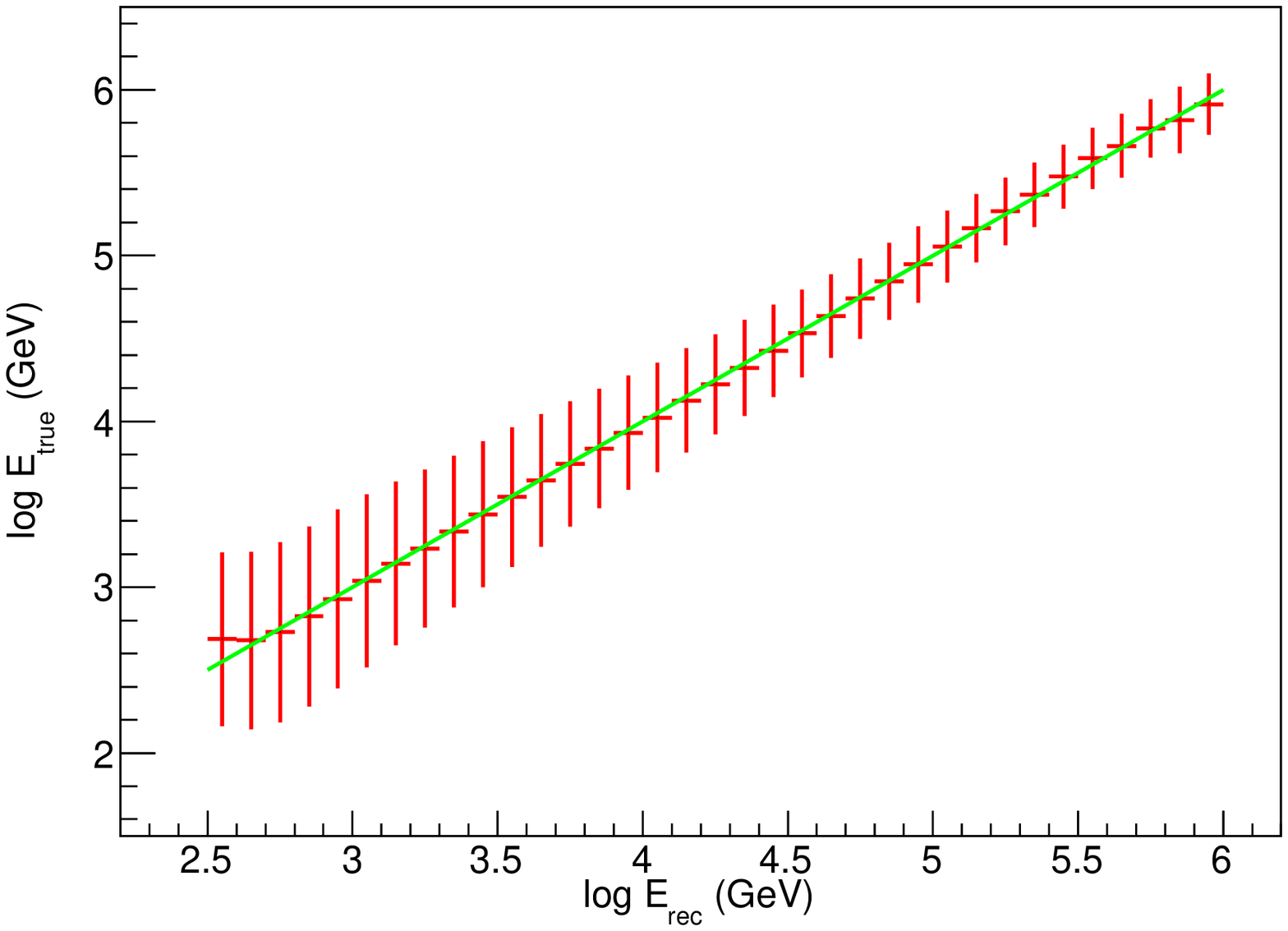}
\caption{Energy reconstruction of the ARGO-YBJ events. Left: The primary energy of cosmic ray ($E_{true}$) as a function of the number of fired strips $N_{str}$. Middle: The difference between $E_{rec}'$ and E$_{true}$ as a function of the secant of the zenith angle $\theta$. Right: The relation between the reconstructed energy and primary energy. The green line indicates a linear function $y=\rm{a}+\rm{b}x$. The error bars along the $x$-axis indicate the width of each bin and those along the $y$-axis indicate the RMS of the parameter distribution.}
\label{energyrec}
\end{figure*}

According to the data selection used in the analysis carried out in \citet{argo2015}, only the data collected in days with   average temperature inside the ARGO Hall $\langle T_{in}\rangle > 10^{\circ}$C are used. This selection removed 35, 78, 130, and 87 days of data during the years 2008, 2010, 2011, and 2012, respectively.

\subsection{Energy reconstruction and data selection}
The cosmic-ray energy has been estimated by a Monte Carlo (MC) simulation. Five groups of dominant component elements, H, He, CNO, Mg-Si, and Fe, are generated according to \citet{Gaisser2013}. The interaction of cosmic rays in the atmosphere is simulated by CORSIKA code v.7.4005 \citep[]{corsika1998}, with the hadronic interaction model GHEISHA at lower energy and QGSJET-II at higher energy. About $4\times10^{10}$ events are sampled with zenith angle distributed from
$0^{\circ}$ to $60^{\circ}$ and energy distributed from 10 GeV to 10 PeV. The detector response is simulated with the G4argo code \citep[]{g4argo} based on GEANT-4.

Generally, the primary energy of an event is positively correlated with the number of fired detectors. The number of fired pads was solely adopted to infer the primary energy in our previous analysis \citep[]{barto13b,argo2015}.
In fact, for events with same number of fired strips (or pads), the true primary energy $E_{true}$ is also related to the incident zenith angle.
In this analysis, the primary energy is first reconstructed based on the number of fired strips $N_{str}$. The left panel of Figure \ref{energyrec} shows the average relation between $N_{str}$ and $E_{true}$. The reconstructed energy at this step is denoted as $E_{rec}'$. Second, the reconstructed energy is further corrected based on the zenith angle. The middle panel of Figure \ref{energyrec} shows the difference between $E_{rec}'$ and $E_{true}$ as a function of zenith angle. The final reconstructed energy is denoted as $E_{rec}$. The right panel of Figure \ref{energyrec} shows the relation between $E_{true}$ and $E_{rec}$, which is a good linear relation.

For the analysis presented in this paper, events are selected according to the following cuts: (1) more than 20 fired strips in the central carpet ($N_{str}\geq 20$), (2) zenith angle less than $50^\circ$, (3) core location less than 100 m from the carpet center, and (4) reconstructed energy $E_{rec}\geq 10^{3.5}$ GeV. Finally, about $3.03 \times 10^{10}$ events survived. The median energy is about 7 TeV.

\subsection{Analysis methods}
The same analysis methods as used in \citet{argo2015} are adopted in this work. The background map is estimated via the equi-zenith angle method based on an iterative procedure. More details about this approach can be found elsewhere \citep[]{AS20051, method2012}. This method can reduce the influence of instrumental and environmental variations. With this approach, the 2D LSA anisotropy can be determined.
The 2D sky in celestial coordinates is divided into a grid of
 $2^{\circ}\times2^{\circ}$ pixels.
The relative intensity in the $(i,j)$th pixel is defined as
\begin{equation}
I_{i,j}=\frac{N_{i,j}}{B_{i,j}}
\end{equation}
where $N_{i,j}$ is the number of detected events, and $B_{i,j}$ is the estimate of isotropic background events.
Due to the Earth's rotation, a ground-based array as ARGO-YBJ is sensitive to the anisotropy in right ascension. The exposure at different declination bands has been estimated with good precision enough to measure the cosmic ray energy spectrum. However, a better accuracy is likely needed to measure a dipolar signal along the declination. Thus this analysis is sensitive to anisotropy in right ascension and the changes in relative intensity across declination belts.

To estimate the amplitude and phase of the anisotropy, a one-dimensional (1D) anisotropy is calculated as the profile of relative intensity in right ascension.
The 1D profile of the anisotropy is fitted by a first-order harmonic function in the form of:
\begin{equation}
I = 1 + A\cos(\alpha-\phi),
\end{equation}
where $A$ is the amplitude of the anisotropy and $\phi$ is the corresponding phase.

\section{Results}

\begin{figure}
\centering
\includegraphics[width=3.3in,height=6.in]{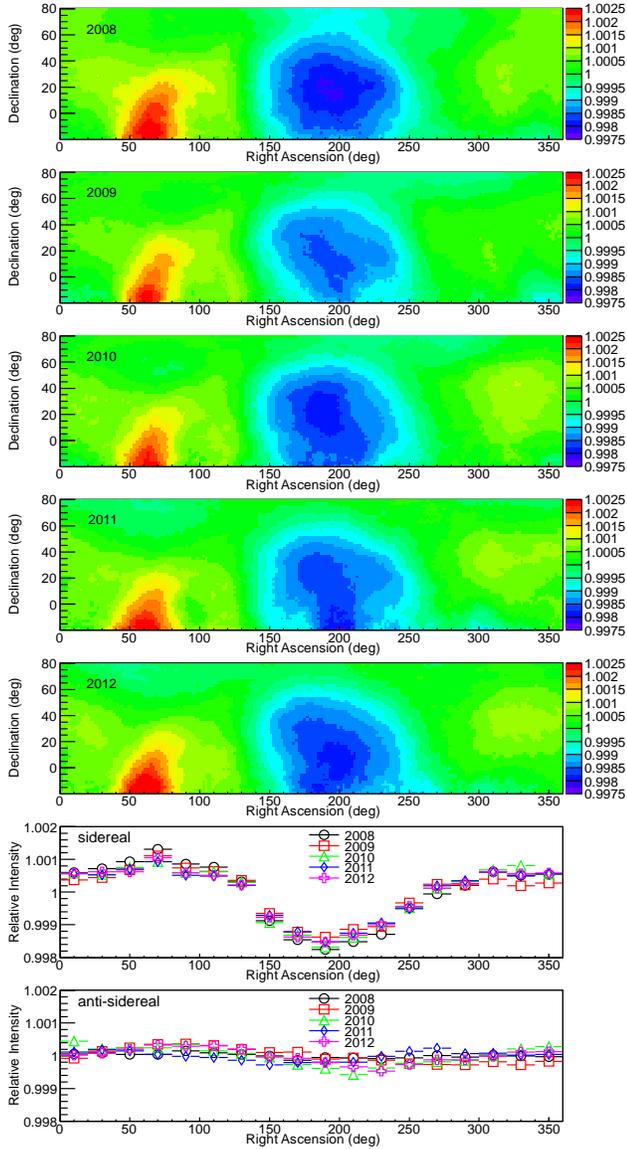}
\caption{The relative intensity of LSA from 2008 to 2012, including all energies. The maps are smoothed with $15^{\circ}$ angular radius. The bottom two panels show the 1D profile of sidereal and anti-sidereal anisotropy, respectively, for years from 2008 to 2012.}
\label{5yrsI2d}
\end{figure}

\subsection{Time dependence of the sidereal anisotropy}
Figure \ref{5yrsI2d} shows the yearly relative intensity of the LSA from 2008 to 2012. The 2D maps are smoothed with an angular radius of $15^{\circ}$. The median energy is about 7 TeV. The "tail-in" and "loss-cone" structures are distinct in the map and almost stable in these years. To check the yearly variation of the anisotropy, the penultimate  panel of Figure \ref{5yrsI2d} shows 1D profiles of the sidereal anisotropy for the five years. The bin size for the 1D profile is $20^{\circ}$ in the direction of right ascension. No substantial or significant variation of the LSA is detected over five years for our total sample with a median energy of 7 TeV.
The anti-sidereal anisotropy, shown in the bottom panel of Figure \ref{5yrsI2d}, should not have a physical origin but rather could arise due to an interaction between a solar diurnal anisotropy and an annual variation.  Such an interaction could also contaminate the sidereal anisotropy.  The bottom panel shows a very weak anti-sidereal anisotropy, which therefore implies very little contamination of the sidereal anisotropy by such an interaction.

The Milagro collaboration reported a steady increase in the amplitude of the "loss-cone" at 6 TeV from July 2000 to July 2007, corresponding to the period from maximum to near minimum of solar cycle 23 \citep[]{milagro2009}. However, the Tibet AS$\gamma$ collabortion reported a stable anisotropy from November 1999 to December 2008 \citep[]{AS2012}.
Located at the northern hemisphere, ARGO-YBJ covers about the same sky region as Milagro and Tibet AS$\gamma$. The energy range of this ARGO-YBJ sample is also very similar to those of Milagro and Tibet AS$\gamma$. The operation period from 2008 to 2012 covers the period of solar cycle from minimum to near maximum. Therefore, ARGO-YBJ is a very suitable detector to address the conflict between those Milagro and Tibet AS$\gamma$ results.

\begin{figure}
\centering
\includegraphics[width=3.3in,height=2.2in]{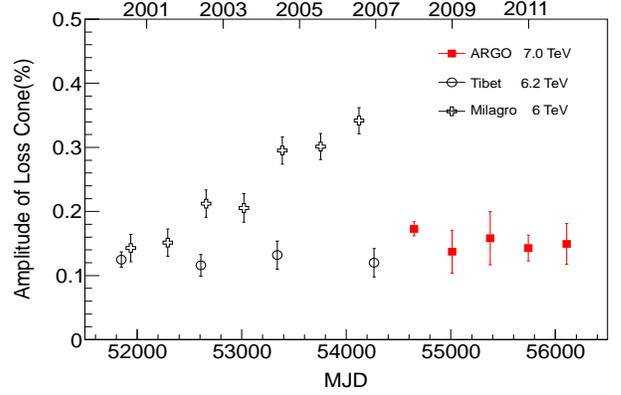}
\caption{Temporal variation of "loss-cone" amplitude of ARGO-YBJ data from January 2008 to December 2012. The error bars represent the sum of systematic and statistical errors. The results of Milagro \citep[]{milagro2009}  and Tibet AS$\gamma$ \citep[]{AS2012} are also presented here for comparison.}
\label{timestable}
\end{figure}

Table \ref{tab0} shows the "loss-cone" amplitude observed by ARGO-YBJ along with the statistical and systematic errors. The "loss-cone" amplitude is defined as the difference between the relative intensity at the minimum of a best fit second-order harmonic function and unity. The statistical error is calculated simply by propagating the statistical errors of the parameters in the fit function. The systematic errors are estimated by the fitted amplitude in anti-sidereal time (the last panel of Figure \ref{5yrsI2d}).
Figure \ref{timestable} shows the yearly "loss-cone" amplitude observed by ARGO-YBJ during the five years. The error bars represent the sum of statistical and systematic errors. The "loss-cone" amplitude is constant within errors. No significant time dependence is observed by ARGO-YBJ during the five years from 2008 to 2012. For comparison, the results of Milagro and Tibet AS$\gamma$ are also shown in Figure \ref{timestable}. Obviously, the ARGO-YBJ result is mostly consistent with the same constant value as the Tibet AS$\gamma$ experiment.
 \begin{table}
 \centering
 \caption{Yearly loss-cone amplitude along with statistical and systematic errors for our total sample with a median energy of 7 TeV}
 \begin{tabular}{cccc}
 \hline\hline
   Year &Loss-cone amplitude (\%) & Stat.error (\%) & Syst.error (\%) \\
   \hline
    2008  & 0.172 & 0.00175 & 0.0076  \\

    2009  & 0.137 & 0.00168 & 0.0300  \\

    2010  & 0.158 & 0.00192 & 0.0376  \\

    2011  & 0.142 & 0.00208 & 0.0160  \\

    2012  & 0.149 & 0.00194 & 0.0279  \\

     \hline\hline
 \end{tabular}
 \label{tab0}
 \end{table}

\subsection{Energy dependence of the sidereal  anisotropy}
\begin{figure*}
\centering
\includegraphics[width=2.3in,height=6in]{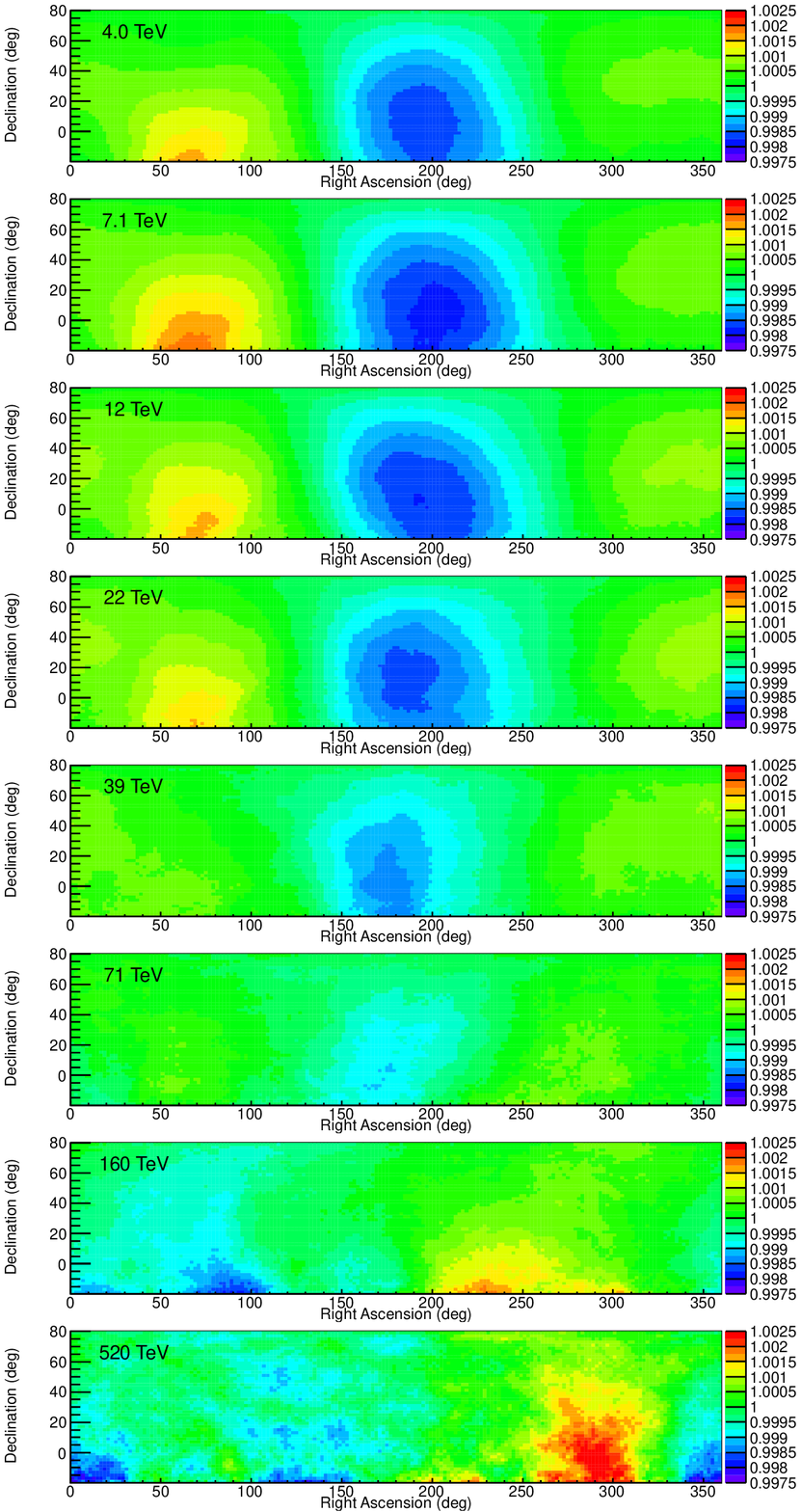}
\includegraphics[width=2.3in,height=6in]{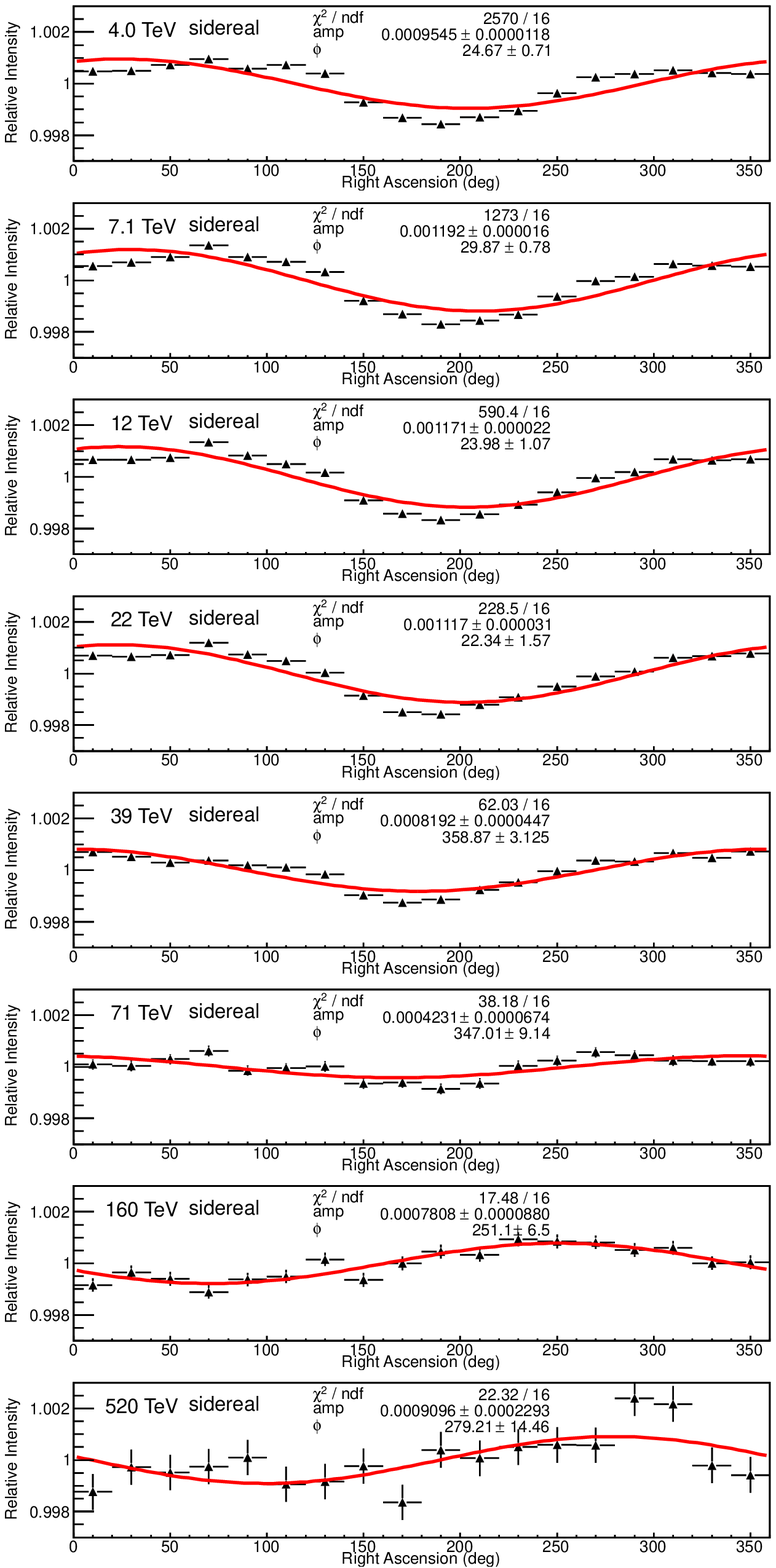}
\includegraphics[width=2.3in,height=6in]{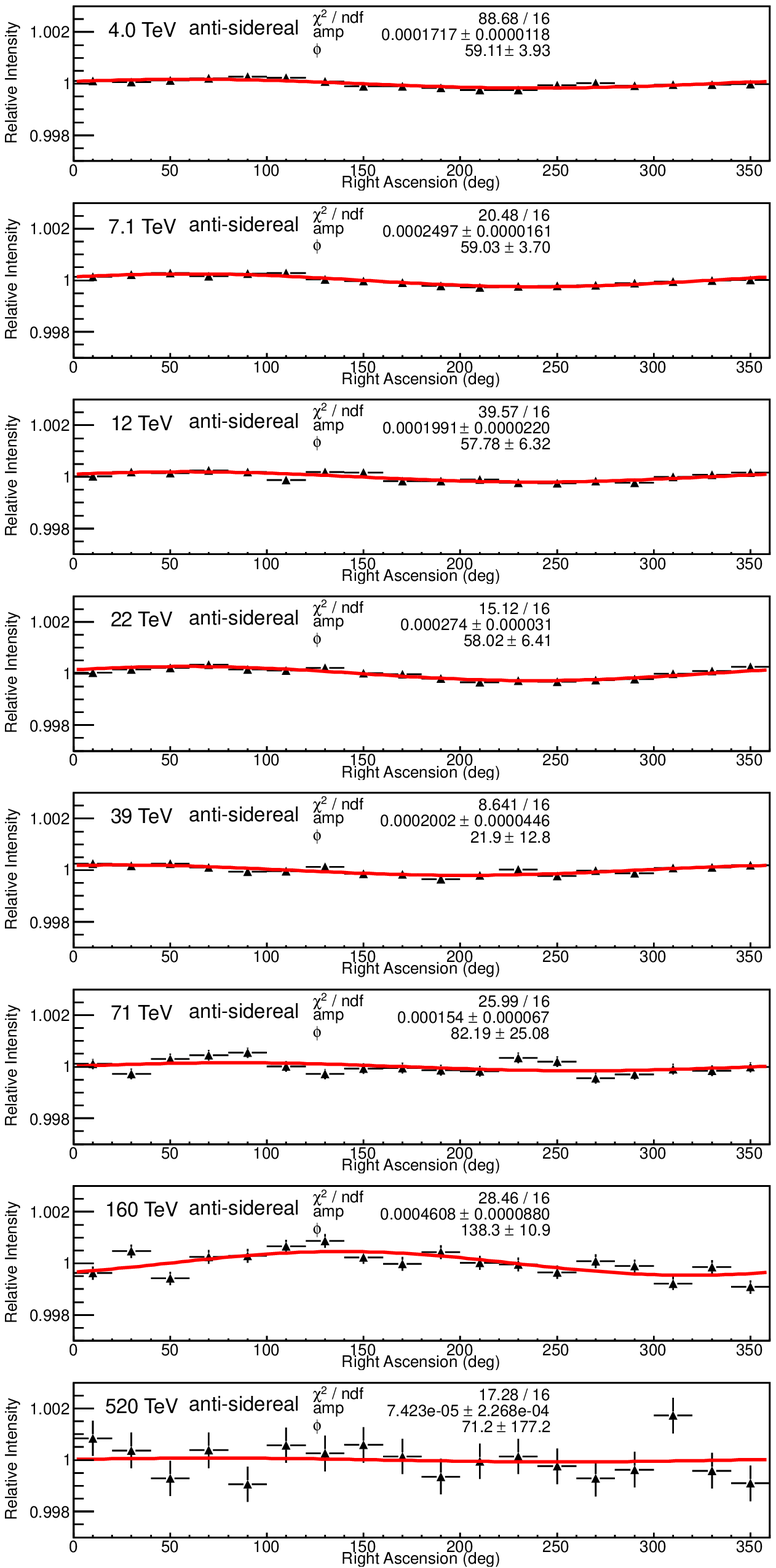}

\caption{The 2D maps of relative intensity of LSA (left panels) and 1D projections on right ascension (middle panels) for 8 energy bins (top to bottom: median energy at 4.0, 7.1, 12, 22, 39, 71, 160, and 520 TeV). Each 2D plot is smoothed with a $30^{\circ}$ angular radius. The median energy of each data sample is labeled at the left. The right panels show the 1D profiles of the anti-sidereal results.}
\label{8E2D1D}
\end{figure*}
To study the energy dependence of the LSA, events are divided into eight samples according to the reconstructed energy $E_{rec}$ as listed in Table \ref{tab1}. The median energies ($E_{m}$) of the primary true energies for the eight intervals are 4.0, 7.1, 12, 22, 39, 71, 160 and 520 TeV, respectively.
It is important to emphasize that the maximum energy, 520 TeV, is much higher than that in our previous analysis, 30 TeV. The energy range presented in this work partly overlaps with that of IceCube \citep[]{icecube2016} and Tibet AS$\gamma$ \citep[]{AS2017}.
The number of events ($N$) in each interval is also listed in Table \ref{tab1}. It is about 1.5$\times10^{10}$ in the first energy interval and gradually decreases to be 4$\times10^{7}$ in the last interval.

 \begin{table}
 \centering
 \caption{Energy bin, median energy, number of events, and amplitude and phase of the LSA dipole component}
 \begin{tabular}{ccccc}
 \hline\hline
   $\log E_{rec}$ (GeV) &$E_{m}$ (TeV) & $N$ ($\times 10^{8}$) & $A$ ($\times 10^{-4}$) & $\phi$ (deg) \\
   \hline
      3.50-3.75 & $4.0$ & 145.0 & 9.54 $\pm$ 0.12 & 24.7 $\pm$ 0.7 \\
      3.75-4.00 & $7.1$ & 77.4 & 11.92 $\pm$ 0.16 & 29.9 $\pm$ 0.8 \\
      4.00-4.25 & $12$ & 41.7 & 11.71 $\pm$ 0.22 & 24.0 $\pm$ 1.1 \\
      4.25-4.50 & $22$ & 21.4 & 11.17 $\pm$ 0.31 & 22.3 $\pm$ 1.6 \\
      4.50-4.75 & $39$ & 10.1 & 8.19 $\pm$ 0.45 & 358.9 $\pm$ 3.1 \\
      4.75-5.00 & $71$ & 4.4 & 4.23 $\pm$ 0.67 & 347.0 $\pm$ 9.1 \\
     5.00-5.50 & $160$ & 2.6 &  7.81 $\pm$ 0.88 & 251.1 $\pm$ 6.5 \\
      $\geq$5.50 & $520$ & 0.4 & 9.10 $\pm$ 2.29 & 279.2 $\pm$ 14.5 \\
     \hline\hline
 \end{tabular}

 \label{tab1}
 \end{table}

Figure \ref{8E2D1D} shows the 2D map of the relative intensity of the LSA for each energy interval. The maps are smoothed with an angular radius of $30^{\circ}$. The "tail-in" and "loss-cone" features are significant at energies from 4 TeV to 22 TeV. At energies from 39 TeV to 71 TeV, the "tail-in" and "loss-cone" features gradually fade away. At the same time, a new excess feature around the right ascension of 250$^{\circ}$-300$^{\circ}$  gradually appears, replacing the structure of the "loss-cone".
At energies above 160 TeV, the "tail-in" and "loss-cone" features completely disappear, and the map is dominated by a new pattern with an excess around the right ascension of 200$^{\circ}$-310$^{\circ}$ and a deficit around 0$^{\circ}$-100$^{\circ}$. These characteristics are consistent with those observed by the Tibet experiment \citep[]{AS2017} in the Northern Hemisphere.

\begin{figure}
\centering
\includegraphics[width=3.3in,height=2.3in]{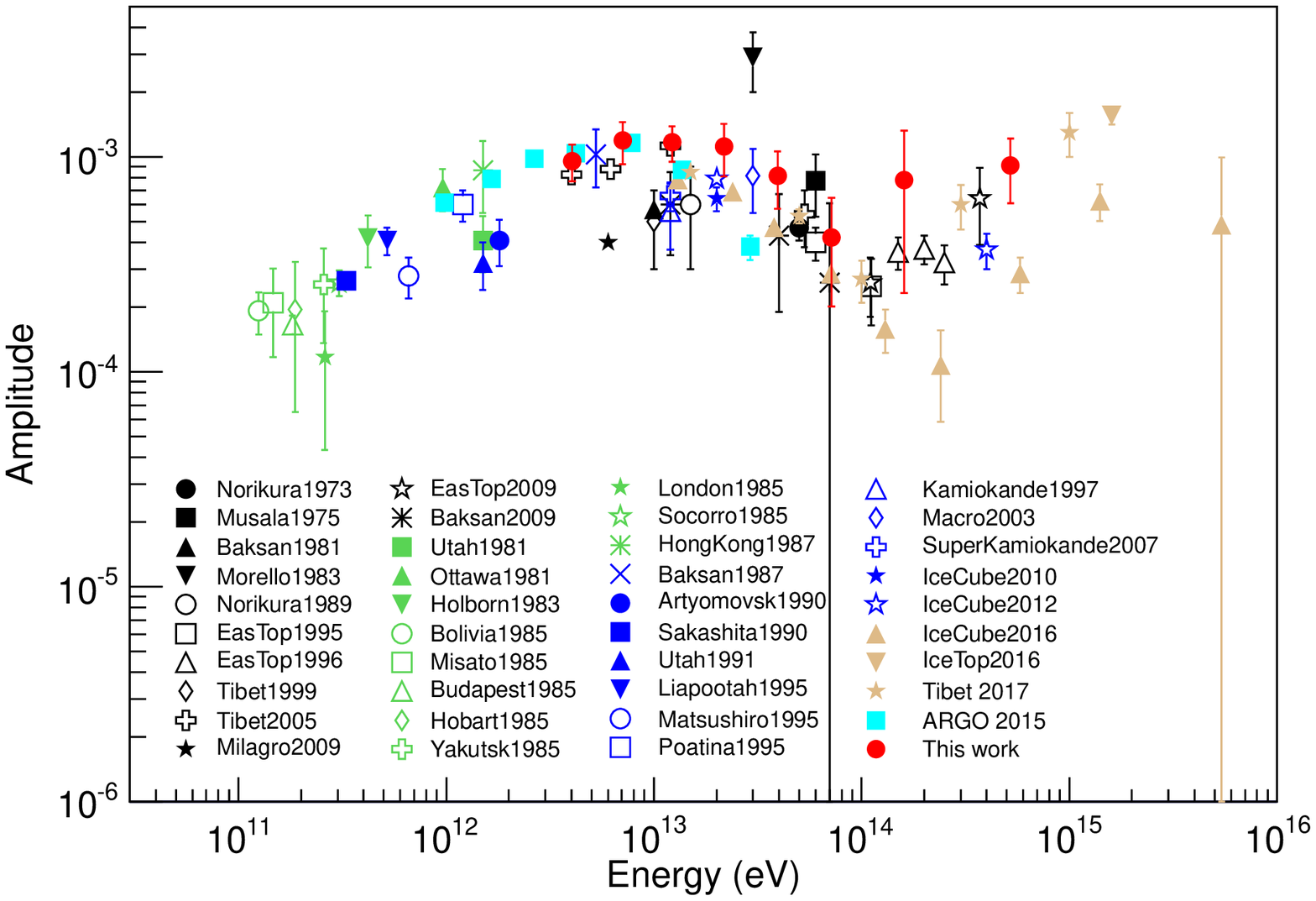}
\includegraphics[width=3.3in,height=2.3in]{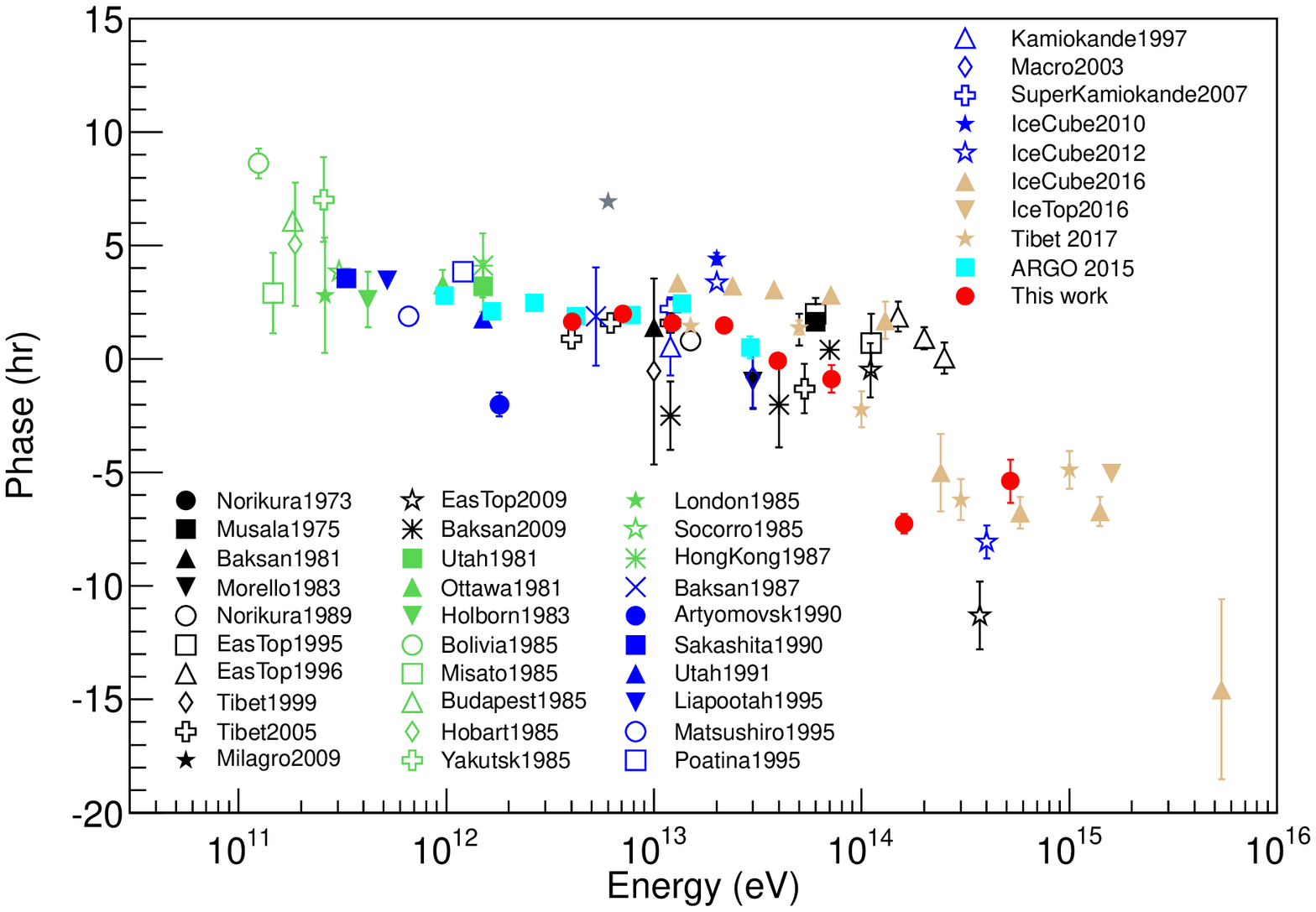}
\caption{The amplitude (top panel) and phase (bottom panel) of the first harmonic of the sidereal anisotropy as a function of the cosmic-ray energy measured by ARGO-YBJ and other experiments \citep[for details and references see][]{Sakakibara1973,Gombosi1975,Alekseenko1981,Cutler1981,Lagage1983,Morello1983,Thambyaphillai1983,Nagashima1985,Swinson1985,Andreyev1987,
Lee1987,Nagashima1989,Kuznetsov1990,Ueno1990,Cutler1991,Aglietta1995,Aglietta1996,Fenton1995,Mori1995,Munakata1997,Munakata1999,Ambrosio2003,
AS2005b,Guillian2007,milagro2008,milagro2009,eastop2009,Alekseenko2009,icecube2010,icecube2012,argo2015,icecube2016,AS2017}.}
\label{APvsE}
\end{figure}

To quantitatively estimate the evolution of the LSA, equation (2) presented in Section 2.3 is used to fit the 1D profiles shown in Figure \ref{8E2D1D} (middle panels). The chi-square values indicate that, mainly at low energies, the cosmic-ray anisotropy is not well described by a pure dipole, but a simple dipole fit is commonly used to estimate its amplitude and phase. The fitted parameters, i.e., the amplitudes and phases, as a function of energy are shown in Figure \ref{APvsE} and also listed in Table \ref{tab1}. The error bars of the amplitude contain the statistical errors and the systematic errors exhibited by the anti-sidereal 1D profiles of Figure \ref{8E2D1D} (right panels). As shown the amplitudes are energy dependent with a maximum around 7 TeV, above which the amplitude begins to decrease with the phase gradually shifting. At energies above 100 TeV, a sudden change of the phase is observed and the amplitude also begins to increase.
According to the fitting parameters shown in Figure \ref{8E2D1D},
the significance of non-zero amplitude is 8.8$\sigma$ and 4.0$\sigma$ at energies of 160 TeV and 520 TeV, respectively, implying that the obtained LSA features are significant.
The phases at 160 TeV and 520 TeV are $\alpha = 251.1^{\circ} \pm 6.5^{\circ}$ and $\alpha = 279.2^{\circ} \pm 14.5^{\circ}$, respectively, which are consistent with Tibet AS$\gamma$ \citep[]{AS2017} and IceCube \citep[]{icecube2016}. The direction of the new excess is very close to the direction of the Galactic Center ($268.4^{\circ}$ R.A.), hinting that this region is the dominant origin of the cosmic rays.
For comparison, the results reported by other detectors are also shown in Figure \ref{APvsE}. The results obtained in this work generally agree with others.

\subsection{The anti-sidereal anisotropy}
Imitating the sidereal and solar time frames, a spurious time frame with 364.242 cycles a year is defined as the anti-sidereal time frame \citep[]{Farley1954}. It is expected to have no physical signal. Therefore, the anti-sidereal anisotropy is usually used to estimate the systematic error of the sidereal anisotropy.
In this work, we also analyze the anti-sidereal anisotropy using the same data used for Sections 3.1 and 3.2. The 1D profiles of the anti-sidereal anisotropy are shown in Figure \ref{5yrsI2d} and Figure \ref{8E2D1D}, respectively, which are used to indicate the systematic error in this work. The maximum systematic error for the yearly sidereal anisotropy shown in Figure \ref{5yrsI2d} is about $0.03\%$, which is much smaller than the LSA amplitude in sidereal time.
The systematic errors from 4.0 TeV to 71 TeV as well as at 520 TeV are also much smaller than the LSA amplitude in sidereal time. At 160 TeV, the systematic error, about $0.04\%$, seems larger than at other energies, while it is still smaller than the observed sidereal anisotropy.
The observed results for the anti-sidereal frequency support the reliability of the observed sidereal anisotropy presented in this work.

\section{Discussion and summary}
In this analysis, only events with reconstructed energy above 3 TeV are used. During five years, no significant time dependence of the anisotropy is detected. However, this result does not exclude the possibility of time-dependent variation of the anisotropy at lower energy. According to the estimation of \citet{zhang2014}, the magnetic field within the heliosphere has minor influence on anisotropy above 4 TeV, and the influence will be visible  at energies below 1 TeV. According to Figure \ref{energyrec}, ARGO-YBJ can also reach the sub-TeV energy band. A study
of the behavior of the anisotropy at energies below 3 TeV  is deferred to a future publication.

This paper reports on the measurement of the large-scale cosmic-ray anisotropy by the ARGO-YBJ experiment with data collected from January 2008 to December 2012. This analysis extends a previous report limited to the period from 2008 January to 2009 December, near the minimum of solar activity between cycles 23 and 24. In contrast with a previous report by the Milagro experiment, no significant time dependence of the anisotropy is detected for a median energy of 7 TeV during 5 years, when the solar activity changed from near minimum to maximum of solar cycle 24. With an improved energy reconstruction procedure, we extended the energy range investigated by ARGO-YBJ up to 520 TeV. A dramatic change of the morphology, consistent with the observations reported by IceCube in the southern hemisphere and Tibet AS$\gamma$ in the northern hemisphere, is clearly observed starting from about 50 TeV. The dipole at 160 TeV and 520 TeV is aligned (at $\alpha=251.1^{\circ}\pm6.5^{\circ}$ and $\alpha=279.2^{\circ}\pm14.5^{\circ}$, respectively) near the direction of the Galactic Center ($268.4^{\circ}$ R.A.), suggesting this region as a possible source of cosmic rays.

\acknowledgments
This work is supported in China by the National Natural Science Foundation of China (NSFC) under the grant No.11575203, No.11375052, No.11635011 and No.11761141001, the Chinese Ministry of Science and Technology, the Chinese Academy of Sciences (CAS), the Key Laboratory of Particle Astrophysics, Institute of High Energy Physics (IHEP), in Italy by the Istituto Nazionale di Fisica Nucleare (INFN), and in Thailand by grant RTA5980003 from the Thailand Research Fund. We also acknowledge essential support from W.\ Y.\ Chen, G.\ Yang, X.\ F.\ Yuan, C.\ Y.\ Zhao, R.\ Assiro, B.\ Biondo, S.\ Bricola, F.\ Budano, A.\ Corvaglia, B.\ D'Aquino, R.\ Esposito, A.\ Innocente, A.\ Mangano, E.\ Pastori, C.\ Pinto, E.\ Reali, F.\ Taurino, and A.\ Zerbini in the installation, debugging, and maintenance of the detector.\\

\end{document}